\documentclass{an}

\usepackage{graphicx,fancyhdr}
\usepackage{rotating}

\usepackage{times}

\begin{document}

\sloppy

\title{Mining the ESO WFI and INT WFC archives for known Near Earth Asteroids. Mega-Precovery software
\thanks{Using ESO/MPG WFI images served by the ESO Science Archive Facility and INT WFC images served by the CASU Astronomical Data Centre. }}

      \author{O. Vaduvescu\inst{1,2,3}\fnmsep\thanks{\email{ovidiuv@ing.iac.es}\newline}, 
              M. Popescu\inst{2,4,7},
              I. Comsa\inst{5},
              A. Paraschiv\inst{9,10},
              D. Lacatus\inst{9,10},
              A. Sonka\inst{6,7},
              A. Tudorica\inst{11,12},
              M. Birlan\inst{2},   
              O. Suciu\inst{5,13},
              F. Char \inst{14},
              M. Constantinescu\inst{7},
              T. Badescu\inst{8,12},
              M. Badea\inst{8,11,12},
              D. Vidican\inst{7},
              C. Opriseanu\inst{7}
              }

\titlerunning{Mining the ESO WFI and INT WFC archives for known NEAs. Mega-Precovery. }
\authorrunning{O. Vaduvescu et al}

   \institute{Isaac Newton Group of Telescopes, 
              Apartado de Correos 321, E-38700 Santa Cruz de la Palma, Canary Islands, Spain 
            \and
              IMCCE, Observatoire de Paris, 
              77 Avenue Denfert-Rochereau, 75014 Paris Cedex, France 
            \and
              Instituto de Astrofisica de Canarias, 
              c/Via Lactea s/n, 38200 La Laguna, Tenerife, Spain 
            \and
              Polytechnic University of Bucharest, Faculty of Applied Sciences, 
              Department of Physics, Bucharest, Romania 
            \and
              Babes-Bolyai University, 
              Faculty of Mathematics and Informatics, 400084 Cluj-Napoca, Romania 
            \and
              Admiral Vasile Urseanu Observatory, 
              B-dul Lascar Catargiu 21, sect 1, Bucharest, Romania 
            \and
              Bucharest Astroclub, 
              B-dul Lascar Catargiu 21, sect 1, Bucharest, Romania 
            \and
              University of Bucharest, Faculty of Physics, 
              Str. Atomistilor 405, 077125 Magurele - Ilfov, Romania
            \and
              Institute of Geodynamics Sabba S. Stefanescu, 
              Jean-Louis Calderon 19-21, Bucharest, Romania, RO-020032, Romania
            \and
              Research Center for Atomic Physics and Astrophysics, Faculty of Physics, University of Bucharest, 
              Atomistilor 405, CP Mg-11, 077125 Magurele - Ilfov, Romania
            \and
              Bonn Cologne Graduate School of Physics and Astronomy, Germany 
            \and
              Rheinische-Friedrich-Wilhelms Universitaet Bonn, Argelander-Institut fur Astronomie, 
              Auf dem Hugel 71 D-53121 Bonn, Germany 
            \and
              The Romanian Society for Meteors and Astronomy, OP 14 OP 1, 130170, Targoviste, Romania 
            \and
              Unidad de Astronomia, Universidad de Antofagasta, Avenida Angamos 601, Antofagasta 1270300, Chile
             }

\date{Accepted for publication in Astronomische Nachrichten (Sep 2012)}

%\keywords{astrometry, minor planets, archives, data mining}

% -------------- Abstract

\keywords{Minor Planets, Near Earth Asteroids, Data Mining, Image Archives, Orbital Amelioration}

\abstract {Abstract: The ESO/MPG WFI and the INT WFC wide field archives comprising 330,000 
           images were mined to search for serendipitous encounters of known Near Earth Asteroids 
           (NEAs) and Potentially Hazardous Asteroids (PHAs). A total of 152 asteroids (44 PHAs 
           and 108 other NEAs) were identified using the PRECOVERY software, their astrometry being 
           measured on 761 images and sent to the Minor Planet Centre. Both recoveries and 
           precoveries were reported, including prolonged orbital arcs for 18 precovered objects 
           and 10 recoveries. We analyze all new opposition data by comparing the orbits fitted 
           before and after including our contributions. We conclude the paper presenting 
           {\it Mega-Precovery}, a new online service focused on data mining of many instrument 
           archives simultaneously for one or a few given asteroids. A total of 28 instrument 
           archives have been made available for mining using this tool, adding together about 
           2.5 million images forming the {\it Mega-Archive}. } 

\maketitle

% -----------------------------------------------------------------

\section{Introduction}
\label{intro}

Telescopes endowed with large field mosaic cameras having their images 
archived and stored for public online access are becoming very appealing 
nowadays to data mining work for many science aims. One of such aim involves 
the improvement of the orbital knowledge of the Near Earth Asteroids (NEAs) 
and Potentially Hazardous Asteroids (PHAs) which is one of the aims of the 
{\it European Near Earth Asteroid Research} (EURONEAR) project since 2006. 

To achieve this goal, few years ago we introduced the PRECOVERY software 
devoted to search {\it all} known asteroids to date in {\it any} archive 
uploaded as a simple observing log recorded in a standard format (\cite{vad09}). 
This tool uses the SkyBoT web service (\cite{ber06}) to predict accurate 
positions for all known asteroids. Using PRECOVERY, all known asteroids could 
be searched in any archive for serenditipitous recoveries and precoveries 
(apparitions before discovery) including all known NEAs, PHAs, as well as 
all other numbered and provisionally named asteroids catalogued to date. 
Using this server, we searched the CFHTLS archive finding 143 encounters 
of known NEAs and PHAs (\cite{vad11a}). 

A recent similar initiative coordinated with the EURONEAR efforts 
includes a citizen-science project led by E. Solano from the Spanish Virtual 
Observatory (\cite{svo11}; \cite{sol11a}). Following its press announcement
(\cite{mun11}), this Spanish service has registered more than 3,000 users who 
measured and reported more than 600 positions of some 150 NEAs (\cite{sol11b}). 
Besides its public outreach value, this work has a meritorious contribution to 
report moving sources not detected by the SDSS automated detection algorithm 
and not published in the SDSS moving source catalogues (\cite{ive08}). The 
above number of detections could be compared with another SDSS NEO project 
which found 104 NEAs in the SDSS archive (\cite{ken09}).

A similar focused tool of NASA was announced recently (\cite{yau11}). 
The {\it Moving Object Search Tool for NEOWISE and IRSA} (MOST) is a new 
web-based server that enables researchers to look for serendipiously observed 
solar system objects (NEAs, asteroids and comets) contained in the images 
held by the NASA/IPAC Infrared Science Archive (IRSA), including single 
epoch exposures from WISE. MOST takes as input an object name or set of 
orbital parameters. 

Part of a student bachelor project, the ESO/MPG WFI archive (1999 to 2003)
was data mined for photometry of known asteroids using the Astro-WISE and 
SkyBoT servers (\cite{bou07}). Taking into account three selection parameters, 
the author measured mostly two colour photometry from 354 occurrences of 144 
asteroids (primarily Main Belt Asteroids - MBAs) on 1380 WFI images. 

Another asteroid data mining tool was announced recently (\cite{gwy11}). 
The {\it Solar System Object Search} (SSOS) of the Canadian Astronomical Data 
Center (CADC) allows users to search for images from a variety of archives for 
{\it single} moving objects, accepting as input either an object designation, 
a list of observations, a set of orbital elements or a user-generated ephemeris 
for an object. 

During the last years we applied our PRECOVERY work to other large field 
archives, also applying our work to other data mining facilities (\cite{vad09}; 
\cite{vad11a}). In the same frame of the EURONEAR project, we introduced 
the {\it Mega-Precovery} project (\cite{pvb10}; \cite{pv10}; \cite{pv11}), 
a web service dedicated to data mining of image archives for some given 
asteroids. Using this tool, one could search one or more existing archives 
for any given asteroid or a list of few asteroids (numbered or provisionally 
numbered). 

In the present paper we present new data mining contributions carried out in 
the frame of EURONEAR project. Firstly, the ESO/MPG WFI and INT WFC archives 
observed between 1998 and 2009 are mined for all known NEAs and PHAs. We 
distributed this workload to a team of 13 Romanian students and amateur astronomers, 
thus the project has some educational aim, besides its main EURONEAR development 
role. Secondly, we present our public new EURONEAR data mining service 
{\it Mega-Precovery}, together with its associated {\it Mega-Archive}. 
Throughout this paper and conform with our two previous papers, we 
define ``precoveries'' as apparitions before discovery date (\cite{ste97}). 

The paper is structured in 5 sections. Section~\ref{datamining} presents the 
ESO/MPG WFI and INT WFC archives, giving an overview of their basic characteristics. 
Section~\ref{results} presents the data mining results, counting the encounters of 
NEAs and PHAs, listing the precoveries and recoveries at a new or the last 
opposition whose orbital improvement is analyzed. 
Section~\ref{MegaPrec} presents {\it Mega-Precovery} software and its associated 
{\it Mega-Archive}. Section~\ref{future} lists the conclusions and a few related 
projects.

%__________________________________________________________________

\section{Data Mining the WFI and WFC Archives}
\label{datamining}

Until the apparition of the SDSS 2.5m survey in 2000 and later the dedication of 
Pan-STARRS 1.8m survey telescope in 2007, the ESO/MPG 2.2m telescope equipped 
with the WFI mosaic camera and the INT 2.5m telescope endowed with the WFC 
mosaic camera have been two of the most powerful 2m class large field facilities 
in the world. Available since 1999 and 1998, respectively, and still operating 
and partially devoted to survey work, these two facilities have given to European 
astronomers access to both hemispheres using mosaic cameras more than half 
degree field each. 

\subsection{ESO/MPG WFI Archive}

The Max Planck Garching (MPG) 2.2m telescope is owned by the European Southern
Observatory (ESO) in La Silla, Chile. In 1999 the Wide Field Imager (WFI) was 
mounted at the Cassegrain $F/5.9$ focus of the ESO/MPG (\cite{baa99}). WFI is a 
mosaic camera consisting of a $2\times4$ CCDs $2K\times4K$ pixels each, covering 
a total field of view of $34^\prime \times 33^\prime$ (0.30 square degrees) with 
a pixel scale of $0.24~^{\prime\prime}$/pix. 

\begin{figure}
\centering
    \mbox{\includegraphics[width=7.5cm]{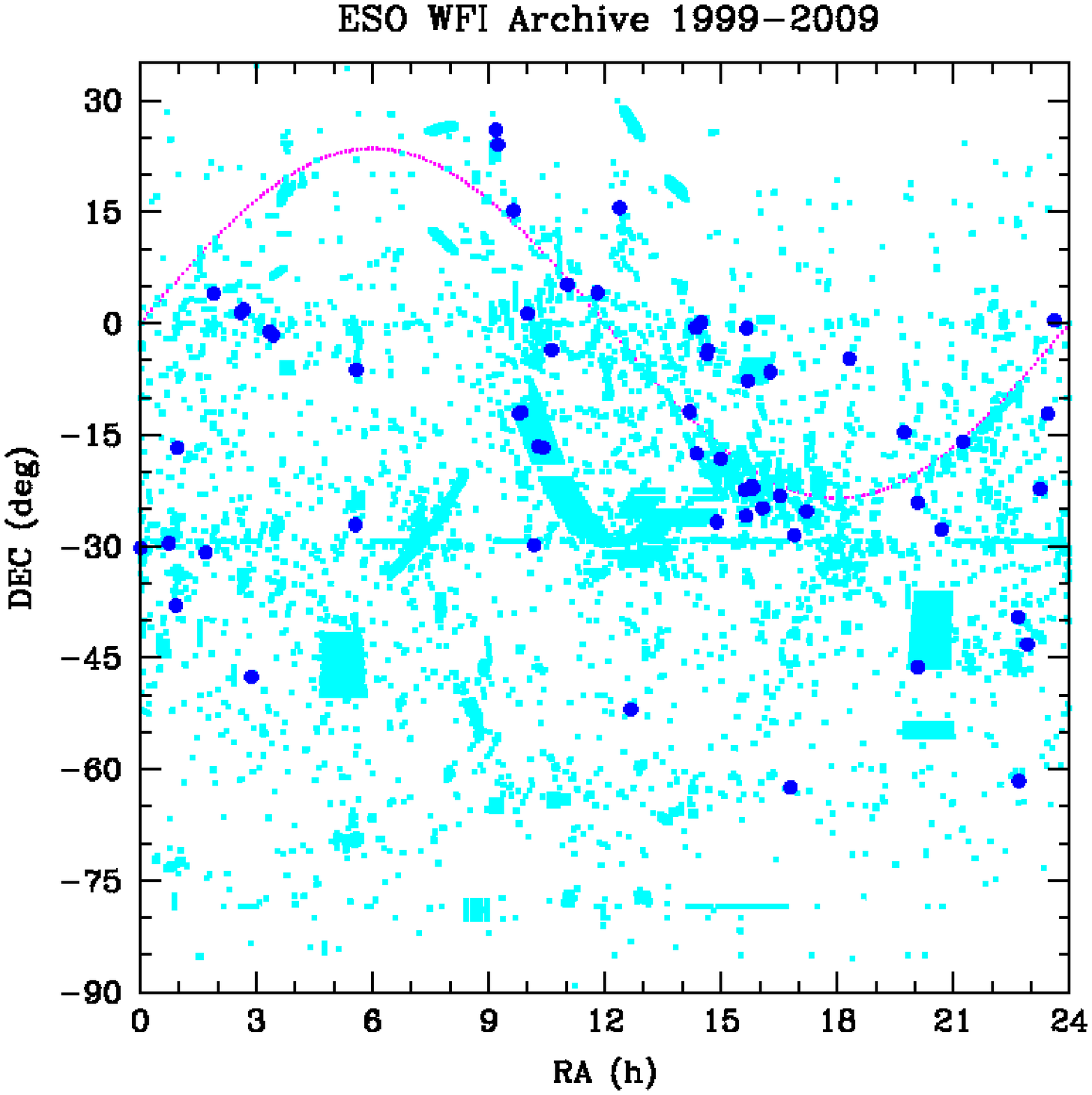}}
    \mbox{\includegraphics[width=7.5cm]{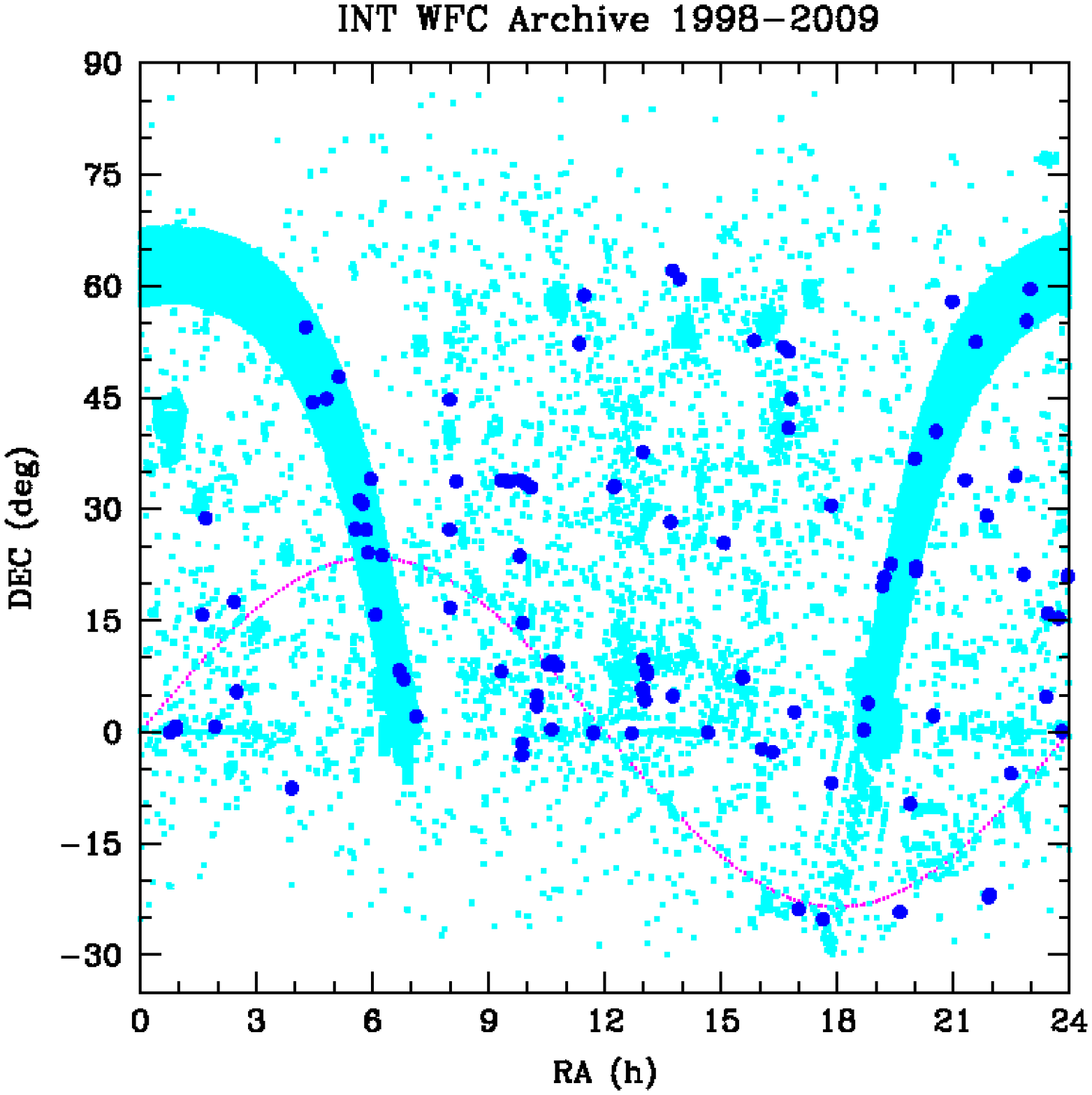}}
\begin{center}
\caption{The sky coverage of the ESO WFI and INT WFC archives whose observed 
fields are plotted as cyan (fainter) dots. Both hemipheres are covered randomly, 
including few sky patches and the galactic plane covered by a few surveys. We 
overlay with blue (larger) dots the NEAs and PHAs encountered in this work and 
with magenta (fainter curve) the ecliptic. 
}
\label{fig1}
\end{center}
\end{figure}

In this paper we studied the ESO/MPG WFI archive during the period 25-10-1999 
(first light) to 27-08-2009 (when we started this ESO project), although WFI has 
continued to be offered by ESO and MPG beyond this date. During this period, the 
WFI acquired 96,913 science images. In Figure~\ref{fig1} (top) we plot in cyan 
(fainter dots) the WFI sky coverage during the above period. This shows random 
pointings between 
$\delta\sim-90^{\circ}$ and $\delta\sim+30^{\circ}$ driven by various science interests 
with some small patches covered by few extragalactic programs. We draw in magenta 
(fainter curve) the ecliptic, which has been followed by at least two NEA survey 
and follow-up programs led by \cite{boa04}; \cite{vad11b} and other Solar System work 
led by other PIs. 

\subsection{INT WFC Archive}

Owned by the Isaac Newton Group (ING), the 2.5m Isaac Newton Telescope (INT) is
installed in the Northern European Observatory of Roque de los Muchachos (ORM) in
La Palma, Canary Islands. Since 1998 the prime focus of the INT houses the Wide 
Field Camera (WFC) which consists of 4 CCDs $2K\times4K$ pixels each, covering an 
L-shape $34^\prime \times 34^\prime$ (0.28 square degrees) with a pixel scale of 
$0.33~^{\prime\prime}$/pix. 

In this paper we studied the INT WFC during the period 20-06-1998 (first light) 
to 10-07-2009 (when we started the INT project), although WFC is continuing to be 
offered by the ING beyond this date. During the above period, the WFC acquired 
237,768 science images. Figure~\ref{fig1} (bottom) plots the sky coverage of 
the WFC archive during this period. The plot shows random distribution between 
$\delta\sim-30^{\circ}$ and $\delta\sim+90^{\circ}$ and clearly some $10^{\circ}$
wide band around the galactic plane which represents the two major galactic 
INT surveys, namely {\it The INT/WFC Photometric $H\alpha$ Survey of the Northern 
Galactic Plane} (IPHAS, \cite{dre05}) and {\it The UV-Excess Survey of the 
Northern Galactic Plane} (UVEX, \cite{gro09}) to be completed soon. 
We draw with magenta the ecliptic, which has been followed sporadically 
by few Solar system runs led by A. Fitzsimmons (NEAs and comets), J.~L. Ortiz 
(TNOs), J. Licandro (MBCs), O. Vaduvescu (NEAs) and possibly other PIs.

\subsection{Exposure Time and Filters}

\begin{figure}
\centering
    \mbox{\includegraphics[width=7.5cm]{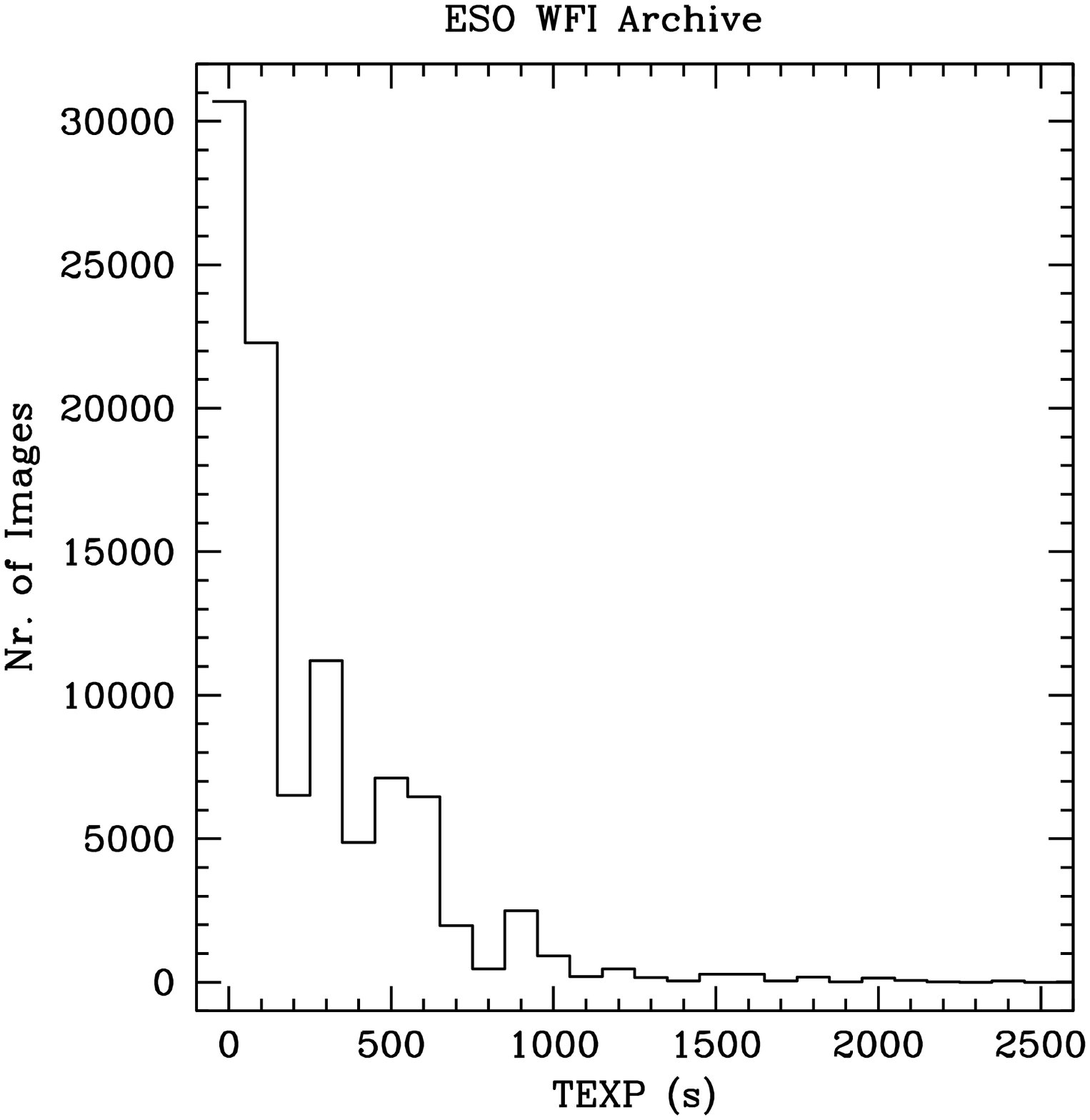}}
    \mbox{\includegraphics[width=7.5cm]{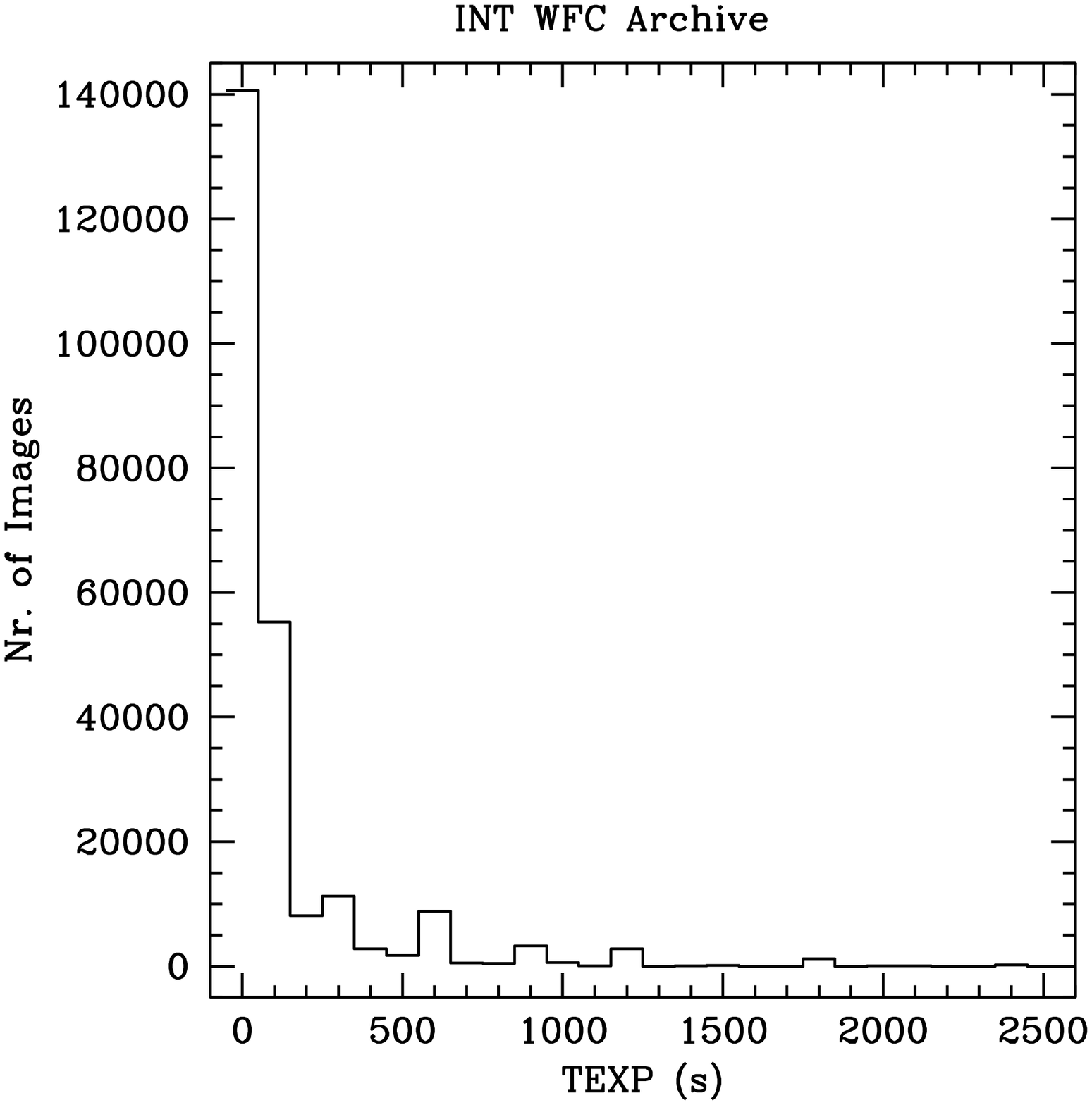}}
\begin{center}
\caption{Histograms showing the exposure time used in the ESO/MPG WFI 
archive (above) and the INT WFC archive (bellow). Relatively short exposures 
(less than 1-2 minutes) were mostly used, making the two archives suitable 
for searching NEAs affected by the trail loss effect. 
}
\label{fig2}
\end{center}
\end{figure}

Figure~\ref{fig2} plots the histogram of the exposure times used in the 
ESO WFI archive (above) and the INT WFC (bellow). Most images were taken with 
exposures shorter than 200s in both archives, therefore they are suitable for 
findings of NEAs affected by a small trail effect caused by their fast proper 
motion ($\mu>1^{\prime\prime}$/min). Few other maxima in exposure times are 
visible on both histograms for longer exposures, up to 2500s in both archives. 

We also studied the usage of filters in the two archives. The ESO WFI archive 
counts 44 filters, while INT WFC counts 20 filters (both broad and narrow band). 
For the ESO WFI archive, $93\%$ represent 13 broad band filters (led by 
$Rc$, $I$, $B$, $V$ and no filter), while 30 other filters representing 
narrow band and others count for only $7\%$. 
For the INT WFC archive, $73\%$ represent 14 broad band filters (led by 
$r$, $i$, $g$, $V$ and no filter) while 6 narrow band filters count for $27\%$. 
In conclusion, both ESO WFI and INT WFC archives are appropriate for asteroid 
detection in most of their broad band images. 

%______________________________________________________________

\section{Results}
\label{results}

\subsection{Found Objects}

Run with the two archives and assuming a safe limiting magnitude $V=23$, 
PRECOVERY reported a total of 7,123 candidate images (1,535 for the ESO WFI 
archive and 5,588 for the INT WFC archive). These images were inspected, 
then astrometrically resolved and measured by our team in a distributed 
but homogenous manner. After inspection and search, only 761 images from 
the initial candidates resulted in reported positions, which represents 
only $11\%$. This dramatic decrease could be explained by some factors, 
led by the optimistic limiting safer magnitude $V=23$, the exposure time, 
proper motion, Moon phase, weather conditions, airmass, uncertainty in some 
ephemerides and magnitudes, used filters, etc. 

In Figure~\ref{fig1} we overlay with blue (larger) dots the NEA and 
PHA fields measured in the two archives (ESO WFC above and INT WFC bellow). 
In both archives the NEA findings are spread randomly with respect to the 
ecliptic up to high ecliptic latitudes ($\beta\sim40^{\circ}$), in agreement 
with the known distribution of the inclinations of the NEA population. 
In total, 152 objects were measured and reported, namely 44 PHAs (15 in 
the ESO and 29 in the INT archive) and 108 other NEAs (40 in the ESO and 
68 in the INT archive). 124 objects have datasets included within the 
timespan of their existing arcs. 
For a total of 28 objects we were able to prolong the existing arcs
by new or last opposition datasets. 
From these, 18 objects represent precoveries (8 in the ESO archive and 
10 objects in the INT one) and 10 objects represent recoveries at a new 
opposition or prolonged arcs at the last opposition (6 in the ESO 
and 4 in the INT archive). 

\begin{figure}
\centering
    \mbox{\includegraphics[width=7.5cm]{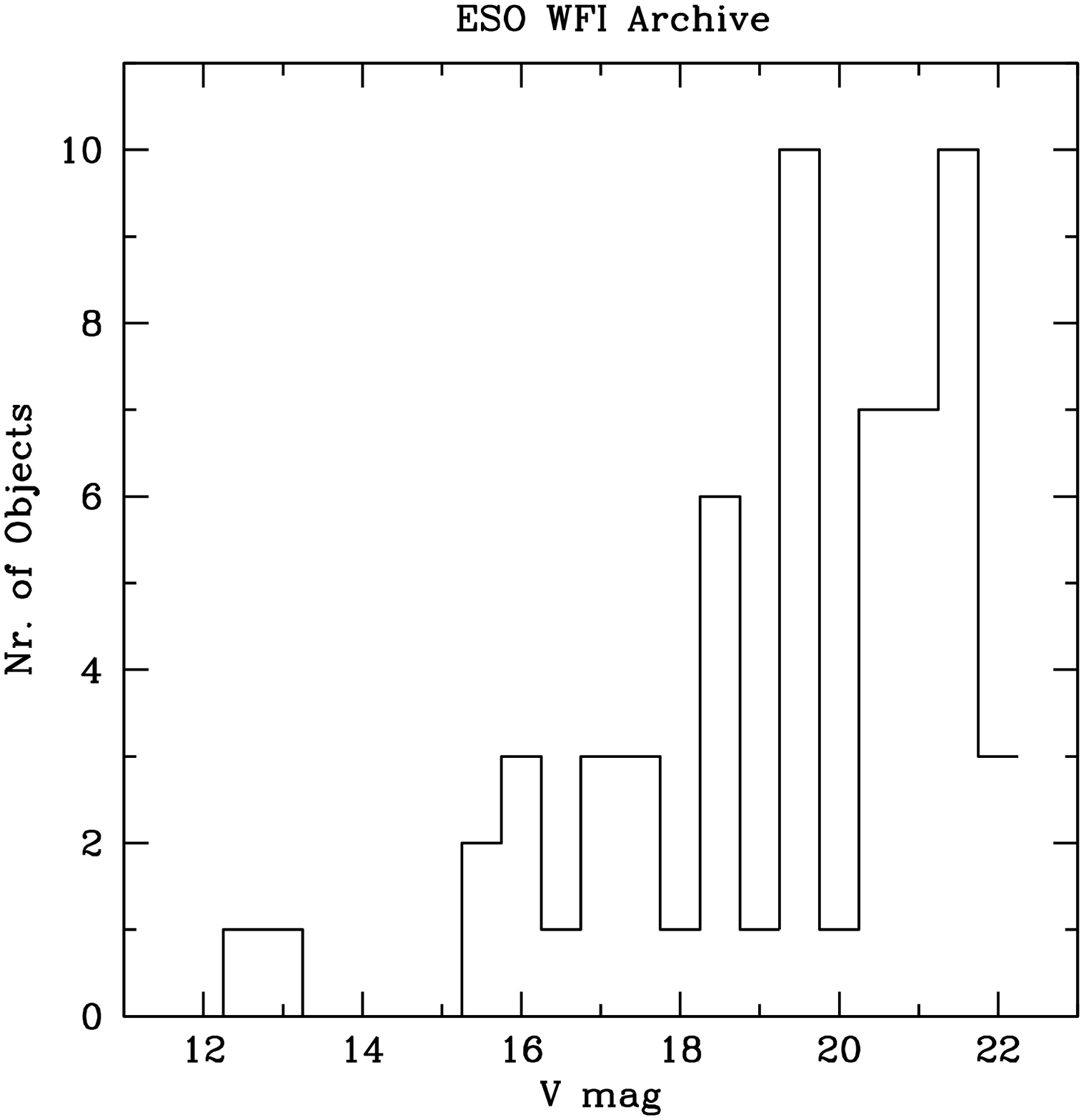}}
    \mbox{\includegraphics[width=7.5cm]{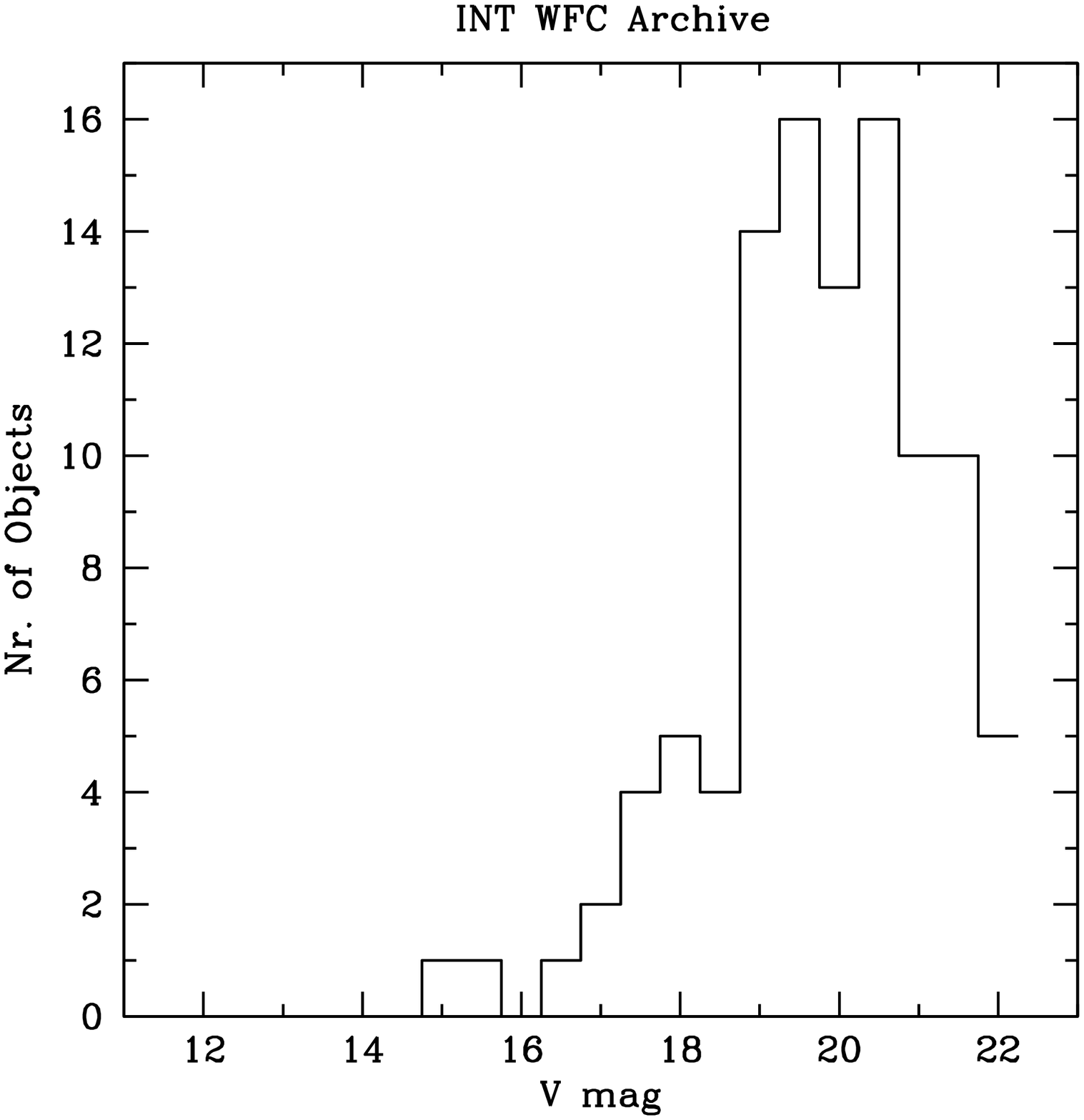}}
\begin{center}
\caption{Histograms showing the predicted $V$ magnitude of the encountered 
NEAs and PHAs in the two surveys (ESO WFC top and INT WFC bottom). A limit 
around $V\sim22$ was reached in both archives, consistent with 2m facilities. 
}
\label{fig3}
\end{center}
\end{figure}

Figure~\ref{fig3} shows the histogram of the predicted $V$ magnitude 
for the encountered objects in the two archives (ESO WFI above and INT 
WFC bellow). Both datasets show fainter objects around $V\sim22$,
according to the limiting magnitude of 2m-class telescopes imaging 
moving asteroids with relatively short exposures (less than $\sim$ one 
minute). 

Using PRECOVERY and {\it Mega-Precovery} (presented in Section~\ref{MegaPrec}), 
one could search the two archives for apparitions of any other given asteroid(s), 
including precoveries of new NEAs and PHAs catalogued after August 2009 (the end 
of our work). We plan to update soon these archives beyound 2009.

\subsection{Astrometry}

Like in our previous distributed work involving students and amateurs, we used 
the {\it Astrometrica} software (\cite{raa12}) to resolve the astrometry of the 
fields. After resolving and aligning the multiple images of the same field, we 
searched for the asteroids in the {\it candidate images} by blinking the images, 
finally measuring manually the asteroid positions. We used quadratic sky-plate 
transformation and the USNO-B1 catalog that allowed in average identification of 
about 50-100 reference stars in each CCD which was resolved individually. 

\begin{figure}
\centering
    \mbox{\includegraphics[width=7.5cm]{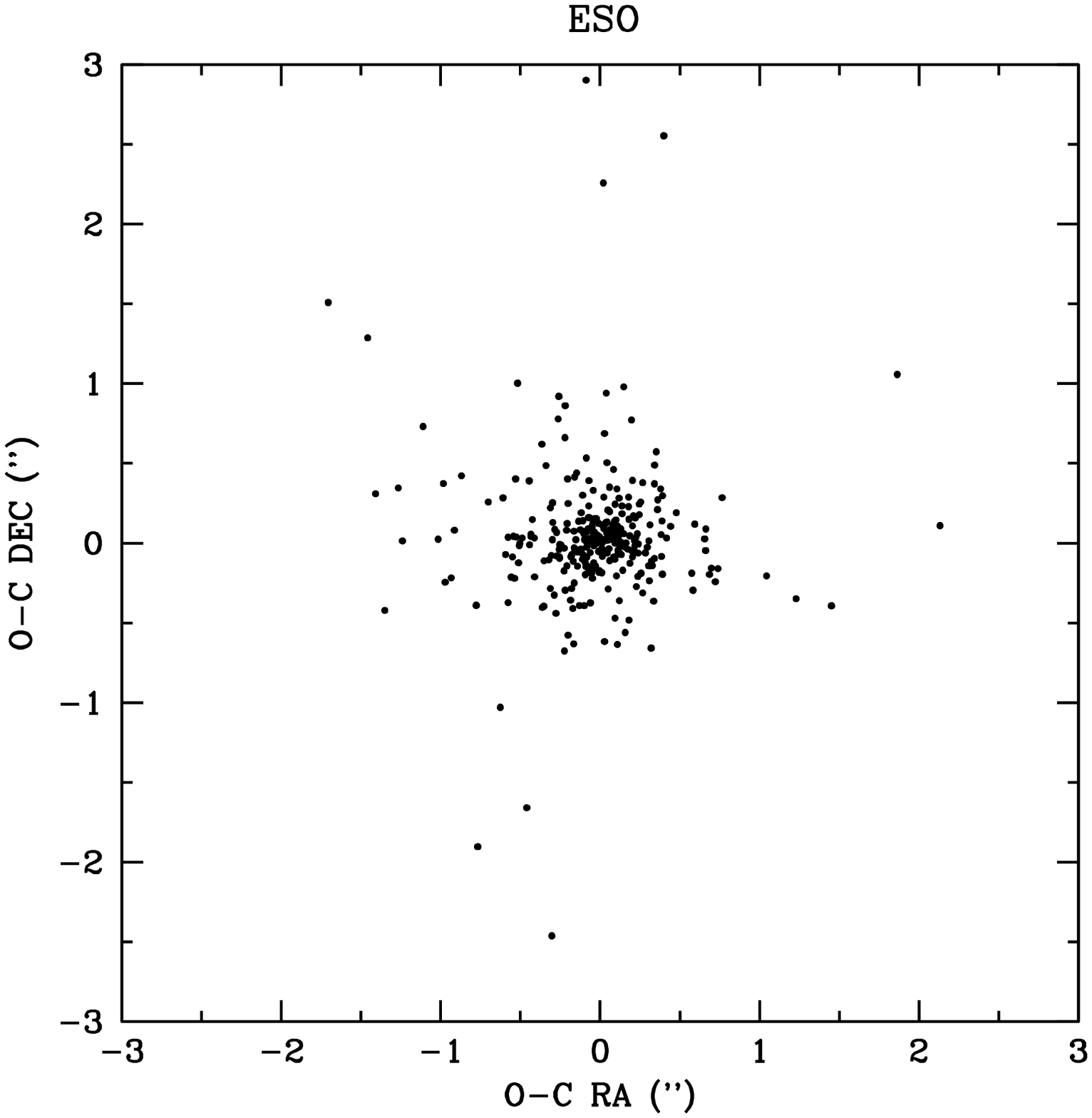}}
    \mbox{\includegraphics[width=7.5cm]{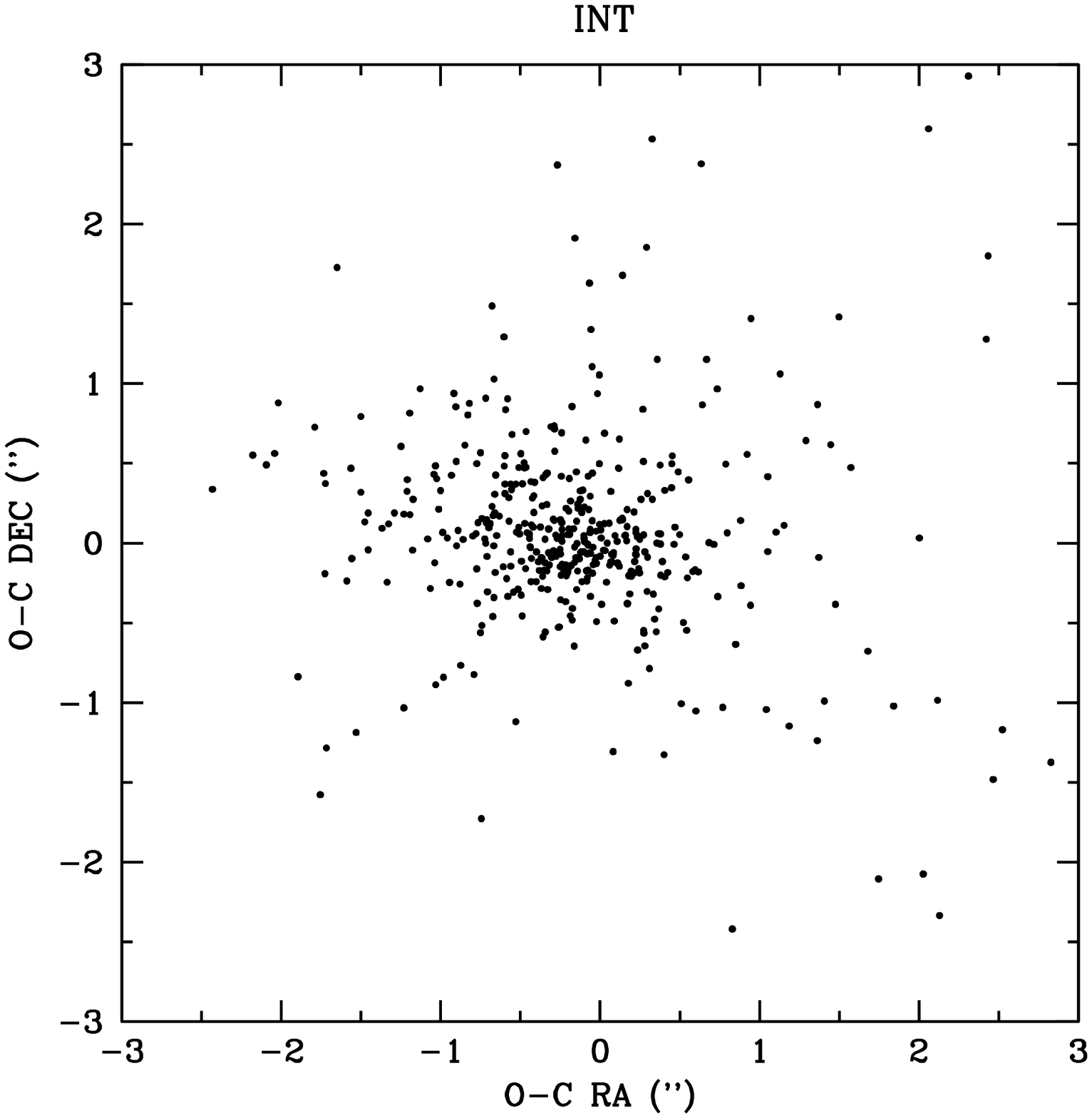}}
\begin{center}
\caption{O-C residuals (observed minus calculated) for the NEAs and PHAs 
         found in the the ESO/MPG WFI (above) and INT WFC archives (bellow). 
         For the ESO WFI, the standard deviation is $0.56^{\prime\prime}$ and the 
         average deviation $0.34^{\prime\prime}$, mostly due by catalog errors. 
         For the INT WFC, the standard deviation is $0.72^{\prime\prime}$ and the 
         average deviation $0.53^{\prime\prime}$, mostly dominated by astrometric 
         errors due to the uncorrected field of the camera mounted in the prime 
         focus of the INT. 
}
\label{fig4}
\end{center}
\end{figure}

Figure~\ref{fig4} plots the O-C residuals (observed minus calculated positions) 
of the objects measured by us in the ESO WFI archive (above) and the INT WFC archive 
(bellow). Residuals are mostly clustered around origin for the ESO WFI archive 
(standard deviation $0.56^{\prime\prime}$ and average deviation $0.34^{\prime\prime}$), 
these being mostly dominated by catalog errors and measurements. The INT WFC 
residuals (right) show a larger spread (standard deviation $0.72^{\prime\prime}$ 
and average deviation $0.53^{\prime\prime}$) mostly due to the larger known WFC 
field distortion in the prime focus of the INT which is larger than that of the 
ESO WFI Cassegrain camera. 

Because most exposure times were quite short (less than one-two minutes) most 
asteroid apparitions show stellar or small elliptical aspect easily fitted by 
Astrometrica which measures their centres in respect to mid-exposure time. We 
encountered also few longer trails caused by longer exposures for which we measured 
in the same manner the two ends, then we measured the average position 
corresponding to the mid-time exposure. 

Using FITSBLINK asteroid residual server calculator (\cite{skv12}) we checked 
the astrometry for possible errors which could include bad measurement of few 
faint targets, faint objects affected by nearby bright stars, possible confusion 
caused by larger sky uncertainties, etc. Then we reported the two datasets 
(one for each archive) to Minor Planet Center (MPC). These include 316 positions 
of 55 objects found in the ESO archive and 445 positions for 97 objects found 
in the INT archive. After minor revision from the MPC (\cite{spa11}) and our 
careful re-measurement of five objects (3\% of the total), MPC accepted and 
published the astrometry (\cite{els11} for the ESO archive and \cite{fit11} 
for the INT archive).

\subsection{Orbital Improvement}

The mined data for the 152 found objects was used to ameliorate orbits 
of the encountered NEAs and PHAs. Most of the data improved the density
within the existing arcs, 
the observing date being contained in the existing orbital arc timeframe. For 
28 objects we prolonged the existing orbital arcs by new oppositions 
or last opposition datasets, adding 18 precoveries and 10 recoveries. 
In the Apendix Table~\ref{table1} we list these 28 cases of NEA and PHA 
encounters representing precoveries and recoveries at a new or the last 
opposition. 
Four of these objects (2005~CG41, 1996~XX14, 2003~MK4 and 2002~DH2) 
were previously reported to MPC in 2005, being observed by other PIs. We mark 
these objects with an asterix in Table~\ref{table1} and Table~\ref{table2}. 
In the third column we list the orbital arc length before and after 
our datamining work. A few cases represent major encounters, namely: 

\begin{itemize}
\item
precoveries at the first opposition: 2005~TU45 (arc prolonged by 5 years), 
2005~MB (extended by 2 years), 2009~TG10 (extended by 6 years), 2009~NA (from 
5 months to 9 years), 2006~PW (from 1 year to 5 years), 2009~HW2 (from 3 
days to 2 months), 2009~SP171 (from 2 months to 2 years), 2009~FU23 (a very 
desirable PHA whose arc was prolonged from 2 months to 7 years) and 2006~KA 
(from 2 years to 6 years); 
\item
recoveries at the last opposition: 1996~XX14 (arc prolonged from 2 months to 
8 years), 2003~MK4 (PHA very desirable, extended arc from 2 months to 2 years), 
2002~DH2 (from 4 months to 3 years) and 1994~JX (arc increased by 5 years). 
\end{itemize}

We compare the orbits of the precovered and recovered objects fitted with 
FIND\_ORB software (\cite{gra12}) using their published positions (full orbital arc) 
available via MPC. In the Appendix Table~\ref{table2} we list the orbital elements 
obtained excluding our data (first line) and including our data (second line). 
Orbits of most asteroids were improved using our mined data, namely the $\sigma$ 
residuals in last column decreased for most. In only 3 cases $\sigma$ residuals 
increased, namely: 
2009~HW2 (from the INT archive, from $\sigma=0.43^{\prime\prime}$ 
to $\sigma=0.48^{\prime\prime}$), extended arc from 3 days to 2 months), 
2001~XK105 (ESO archive, from $\sigma=0.61^{\prime\prime}$ 
to $\sigma=0.63^{\prime\prime}$, for which only two images were encountered), and 
2006~KA (ESO archive, from $\sigma=0.49^{\prime\prime}$ to 
$\sigma=0.53^{\prime\prime}$, for which only one image was available showing a 
long trail for which we reported the middle and the two ends taking into account 
the predicted proper motion). 

%______________________________________________________________

\section{Mega-Precovery}
\label{MegaPrec}

Despite some recent data mining efforts, the vast collection of CCD 
images and photographic plate archives still remains insufficiently exploited. 
PRECOVERY covers all catalogued asteroids (including all known NEAs and PHAs), 
but the search of an entire archive could take quite a long time, typically 
about 10 hours for some 10,000 square degrees sky coverage of a single archive. 
Therefore, some dedicated tool to target one or very few selected objects is 
necessary to speed up data mining of asteroids, including important NEAs and PHAs. 

In this sense we designed {\it Mega-Precovery}, with the aim to fasten and 
target the search of one or some few important objects, such as PHAs or 
Virtual Impactors (VIs). Given this, we propose to search very large {\it 
collections of archives} for images which include one or a few selected known 
asteroids in their field. There are two components of this project: 

\begin{itemize}
\item the database (named {\it Mega-Archive}) which includes the individual 
{\it instrument archives}, namely the observing logs for their science CCD images 
or plates available from a collection of instruments and telescope around the globe. 
The {\it Mega-Archive} is an open project allowing other {\it instrument archives} 
to be added later for exploration by anybody who would like to contribute; 
\item the {\it Mega-Precovery} software for data mining the {\it Mega-Archive} 
for the images containing one or a more desired catalogued objects (NEAs, PHAs
or other asteroids) included in a local daily updated MPC database; 
\end{itemize}

The input of {\it Mega-Precovery} consists of: 1. a selection of the {\it instrument 
archives} to be searched (including the option to search either one or all the existing 
archives in the same time) and 2. the specified asteroid or list of few asteroids given 
by their names, numbers or provisional designations. The output of {\it Mega-Precovery} 
consists of a list of {\it candidate images} in which the searched object is expected 
to be visible based on two main thresholds, namely the expected limiting magnitude 
of the archive and the expected positional uncertainty of the searched object 
(provided by the user based on the currently known orbit). 

The {\it definition file} containing the {\it instrument archives} includes the 
following data: the filename keeping the telescope observing logs, the observatory 
code, the width and height of the field (both in degrees, in the direction of 
$\alpha$ and $\delta$, respectively), the start and end Julian Date defining the 
timespan of the archive and the limiting magnitude $V$ expected from the given 
telescope and instrument. 

\subsection{Mega-Precovery Archive}
\label{archive}

For easier identification of the images, the {\it Mega-Archive} is split into 
more {\it instrument archives}, each corresponding to a given telescope and camera. 
Table~\ref{table3} lists the available {\it instrument archives} and their basic 
characteristics. Besides these standard archives, {\it Mega-Precovery} leaves the
user the flexibility to add his/her own instrument archive given in the same
standard format, by loading the file to be run by the server. As of August 2012, 
the {\it Mega-Archive} counts about 2.5 million images from 28 instrument archives 
available for search via {\it Mega-Precovery}. This collection includes all archived ESO 
imaging instruments\footnote{http://archive.eso.org/eso/eso\_archive\_main.html}, 
most archived NVO imaging instruments (National Virtual Observatory of the 
United States\footnote{http://portal-nvo.noao.edu}), 
the INT WFC\footnote{http://casu.ast.cam.ac.uk/casuadc/archives/ingarch/@@query.html}, 
CFHTLS\footnote{http://www3.cadc-ccda.hia-iha.nrc-cnrc.gc.ca/cadcbin/cfht/wdbi.cgi/cfht/quick/form}, 
Subaru SuprimeCam\footnote{http://smoka.nao.ac.jp/search.jsp}
and the AAT WFI\footnote{http://apm5.ast.cam.ac.uk/arc-bin/wdb/aat\_database/observation\_log/make} 
archives. 

\subsection{Mega-Precovery Software}
\label{software}

\begin{sidewaysfigure}

\centering
    \mbox{\includegraphics[width=15cm]{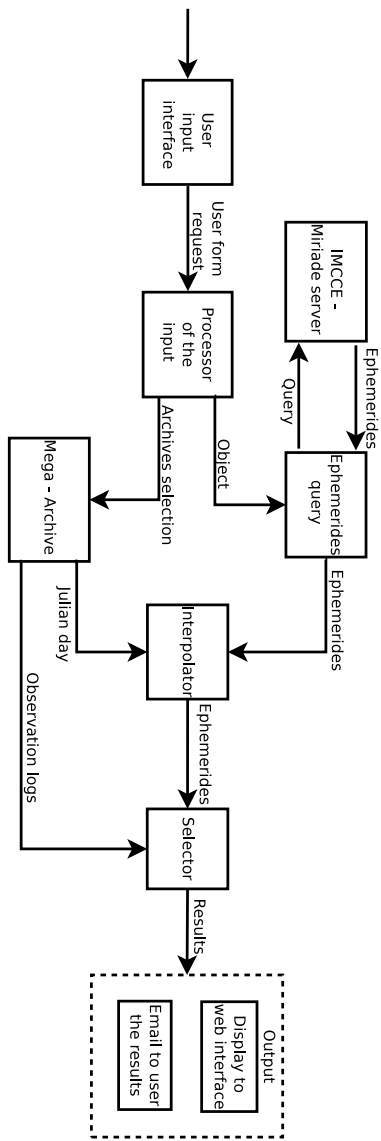}}
\begin{center}
\caption{The flowchart of {\it Mega-Precovery} software. }
\label{fig5}
\end{center}
%\end{figure}
\end{sidewaysfigure}

The {\it Mega-Precovery} software is written in PHP, being embedded on the
EURONEAR website (\cite{eur12}) as a public access application under the 
Observing Tools section. Figure~\ref{fig5} presents the flowchart of this 
software. To run {\it Mega-Precovery} application, the user needs to 
load the webpage using any internet browser (block {\it User input interface} 
in Figure~\ref{fig5}). In order to create the query, one needs to provide 
the following information to the web interface: 

\begin{itemize}
\item A list of names, numbers or provisional designations of the asteroid(s) 
to search; 
\item The selection of the {\it instrument archives} to be searched for (the first 
default option {\it ALL} allowing to search the entire {\it Mega-Archive}); 
\item The field {\it Uncertainty} used to accommodate for the uncertainty in 
telescope pointing plus the uncertainty in position of the searched object due 
to its (sometime more unsecure) orbit. Based on extended tests, for this parameter 
we recommend the default value $0.02^{\circ}$ but this should be increased in case 
of poorly known objects or less accurate pointing. Increasing too much this 
parameter (more than $\sim1^{\circ}$) would result in selection of false candidate 
images (false detections); 
\item The email address where {\it Mega-Precovery} could announce the user about 
the end of long runs and the FTP space where the user should download the data; 
this includes the same information given in the web browser after the end of the run. 
\end{itemize}

After the user submits the query, this is processed by {\it Mega-Precovery} 
in the block {\it Processor of the input}, then the accurate ephemerides 
for each archive dates and the given body (bodies) are calculated in the 
{\it Ephemerides query block}. This step uses the IMCCE's ephemeris service 
{\it Miriade} (\cite{ber09}; \cite{imc12a}) which is queried for some discrete times 
covering the entire queried telescope archive(s), then accurately interpolated 
the positions for the observing dates given in the archive (the block 
{\it Interpolator}). 

We used a five order Bessel interpolation model (\cite{imc12b}), 
choosing for the {\it Miriade} ephemerides a step based on the asteroid 
proper motion, namely 1 day for objects moving slower than 2 degrees daily 
and 1 hr for objects faster than this limit. Using this fine time step we 
ensure sufficient accuracy 
($<1^\prime$) for most NEAs and PHAs passing close to Earth affected by 
very fast proper motion. For NEAs and PHAs away from close encounters with 
the Earth, and also for MBAs, the interpolator precision is very accurate 
(about $1^{\prime\prime}$). The parameters for the interpolation where 
established as a tradeoff between processing time and predicted position 
accuraccy and were validated by extensive tests. 

Each image of the archive is defined by a rectangular box given by the 
telescope pointing, the width and height field of view stored with each 
instrument archive entry. If the predicted position falls within the field 
of the current image bounded by the {\it Uncertainty} area, then the current 
entry is selected as a {\it candidate image} to hold the queried asteroid. 
This step is done in the {\it Selector} block. 

Like in PRECOVERY, the format of each instrument archives follows the same 
{\it standard format} listing one observation (telescope pointing) on each 
line which includes the image ID (name of the image file), the Julian date 
(start of observation), exposure time (sec), telescope pointing ($\alpha$ 
and $\delta$ at J2000 epoch), width and height of the field of view (towards 
$\alpha$ and $\delta$ axis frame, both in degrees), and a eventually comment 
(which could include the filter, etc), all separated by ''$|$''. 

The output of {\it Mega-Precovery} consists in a list including the images 
and the corresponding CCD number predicted to contain the queried object(s). 
The results are displayed both in the web interface (visible only at the 
end of the run) and sent via e-mail to the user (in case this option was 
selected). The user can search the images in the online instrumental archive, 
then download, inspect and measure the data related to this asteroid based 
on his/her own scientific interest (astrometry, photometry, etc). 

%\begin{figure}

%______________________________________________________________

\section{Conclusions and Future Work}
\label{future}

Two wide field 2m class telescope archives, ESO/MPG WFI and INT WFC, 
comprising about 330,000 images were mined to search for serendipitous 
encounters of known NEAs and PHAs. 
Two master archives were built based on the observing logs of the two facilities. 
Using the PRECOVERY software, a total of 152 asteroids (44 PHAs and 108 other NEAs) 
were identified and measured on 761 images and their astrometry was reported to Minor 
Planet Centre (MPC). Both recoveries and precoveries were reported, including 
prolonged orbital arcs for 18 precovered asteroids and 10 recoveries, plus other 
124 recoveries. We analyze all precoveries and recoveries at a new or last 
opposition by comparing the orbits fitted before and after including our datasets. 
Following the PRECOVERY project, we present {\it Mega-Precovery}, a new search 
engine focused on data mining of many instrument archives simultaneously for one 
or few given asteroids. A total of 28 instrument archives have been made available 
for mining, adding together about 2.5 million images forming the {\it Mega-Archive}. 

Few other projects are in plan within the frame of EURONEAR data mining of NEAs. 
Few months ago we started datamining of the Subaru SuprimeCam archive using PRECOVERY. 
Another important project will be to extend the mining capabilities of the 
{\it Mega-Precovery}. 
Another project plans to apply {\it Mega-Precovery} to search the entire 
{\it Mega-Archive} in order to recover and improve orbits of some important VIs, 
PHAs and NEAs. 
Finally, we plan to continue to enlarge the {\it Mega-Archive}, adding new 
{\it instrument archives}, including CCD cameras and photographic plates. 
In this sense, any observatories, especially those endowed with large field 
imaging instruments, are welcomed to contribute to this open source project. 

%______________________________________________________________

\smallskip
\begin{acknowledgements}
This project includes observations made with the ESO/MPG telescope of European 
Southern Observatory (ESO) in La Silla Observatory, Chile and the Isaac Newton 
Telescope (INT) owned by the Isaac Newton Group (ING) in La Palma, Canary Islands. 
This work used images retrieved from the ESO Science Archive Facility and from the 
Isaac Newton Group Archive maintained by the CASU Astronomical Data Centre at the 
Institute of Astronomy, Cambridge. 
We acknowledge Jerome Berthier for his continuous support provided with the 
SkyBoT (\cite{ber06}) and Miriade (\cite{ber09}) servers developed at IMCCE and 
accessed by PRECOVERY and {\it Mega-Precovery}, respectively. 
Acknowledgements are also due to Bill Gray, the author of FIND\_ORB software, for 
his very prompt assistance in order to upgrade his very robust and user friendly 
code for fitting orbits. 
Mega-Precovery archive is based on data obtained from the ESO Science Archive 
Facility, the NOAO Science Archive served by the National Virtual Observatory of 
the United States, Subaru-Mitaka-Okayama-Kiso Archive (SMOKA), the Canadian 
Astronomy Data Centre (CADC) and CASU Astronomical Data Centre (United Kingdom). 
This work has made use of SAOImage DS9 developed by Smithsonian Astrophysical Observatory. 
We are thankful to Minor Planet Centre, specifically to Tim Spahr who pointed out a 
few errors in the reported positions. Final acknowledgements are due to the 
referee David Asher for careful reading and advice to improve our paper. 
\end{acknowledgements}

\appendix
\section{Appendix: Data Tables}

\renewcommand{\arraystretch}{0.8}
\begin{table*}[!t]
\begin{center}	
\caption{28 objects found in the ESO and INT archives include 18 precovered asteroids 
and 10 recovered objects whose arcs were prolonged by a new opposition or at the 
last opposition. Besides the asteroid name we give its MPC classification (acc. 
to Jan 2012 database), the number of positions, the length of the orbital arc (before 
and after our astrometry) and the archive. Four objects marked with an asterix were 
reported also by other PIs in 2005. } 
\label{table1}
\begin{tabular}{llrlr}
\hline
\hline
\noalign{\smallskip}
\noalign{\smallskip}
Asteroid  &  Classification  &  Nr. pos.  &   Arc (before/after)  &  Archive \\
\noalign{\smallskip}\noalign{\smallskip}

\hline
\noalign{\smallskip}\noalign{\smallskip}
\multicolumn{5}{c}{Extended Arcs at First Opposition (Precoveries):} \\
\noalign{\smallskip}

2005 CG41 *        & NEA desirable           & 12 & 3d/4d   & ESO \\
2005 TU45 (231134) & NEA desirable           &  4 & 3y/8y   & ESO \\
2005 MB            & NEA very desirable      &  3 & 3y/5y   & ESO \\
2004 XP164 (216707)& NEA desirable           &  5 & 4y/5y   & ESO \\
2009 TG10          & NEA very desirable      &  6 & 3y/9y   & ESO \\
2009 NA            & NEA very desirable      &  3 & 5m/9y   & ESO \\
2002 HQ11 (159677) & NEA desirable           &  2 & 5y/6y   & ESO \\
2001 XK105         & NEA extremely desirable &  2 & 2m+10d  & ESO \\
\noalign{\smallskip}
\noalign{\smallskip}
2006 WT1           & PHA very desirable      &  6 & 4m/5m   & INT \\
2002 TY57 (250162) & NEA desirable           &  7 & 5y+2m   & INT \\
2006 PW            & NEA very desirable      &  2 & 1y/5y   & INT \\
2009 HW2           & NEA extremely desirable &  8 & 3d/2m   & INT \\
2009 SP171         & NEA very desirable      &  2 & 2m/2y   & INT \\
2005 BY1           & NEA very desirable      & 12 & 10m/11m & INT \\
2009 FU23          & PHA very desirable      &  2 & 2m/7y   & INT \\
2005 VC2           & NEA very desirable      &  5 & 3m/4m   & INT \\
2006 KA            & NEA very desirable      &  3 & 2y/6y   & INT \\
2000 FL10 (86666)  & NEA desirable           &  3 & 8y/9y   & INT \\
\noalign{\smallskip}\noalign{\smallskip}

\hline
\noalign{\smallskip}\noalign{\smallskip}
\multicolumn{5}{c}{Extended Arcs at Last Opposition (Recoveries):} \\
\noalign{\smallskip}

1996 XX14 *        & NEA very desirable     &  7 & 2m/8y    & ESO \\
2003 MK4 *         & PHA very desirable     &  7 & 2m/2y    & ESO \\
2002 DH2 *         & NEA very desirable     &  8 & 4m/3y    & ESO \\
1994 JX            & NEA very desirable     &  8 & 9y/14y   & ESO \\
2004 RS25          & NEA very desirable     &  2 & 3y+1m    & ESO \\
1999 WK11 (102873) & NEA desirable          &  6 & 28y+1m   & ESO \\
\noalign{\smallskip}
\noalign{\smallskip}
2005 GN59 (164400) & PHA desirable          &  2 & 31y+3m   & INT \\
2002 EQ9 (163191)  & NEA very desirable     &  2 & 5y+10d   & INT \\
2007 DK            & NEA very desirable     &  4 & 5y+7m    & INT \\
2005 CA (189263)   & NEA desirable          &  4 & 30y+4d   & INT \\
\noalign{\smallskip}

\hline
\hline
\end{tabular}
\end{center}
\end{table*}

% ________________

\begin{table*}[!t]
\begin{center}
\caption{Extended arcs of NEAs and PHAs at first opposition (precoveries) and last opposition (recoveries). 
         Comparison of the orbits fitted without our data (first line) and including our data (second line). 
         Orbital elements fitted with FIND\_ORB at epoch $JD=2456000.5$ listing the asteroid name, 
         semimajor axis $a$, eccentricity $e$, inclination $i$, longitude of the ascending node $\Omega$, 
         argument of pericenter $\omega$, and mean anomaly $M$, followed by the minimal orbital intersection 
         distance MOID, number of fitted observations $Obs$ and the root mean square residual of the fit 
         $\sigma$. Four objects marked with an asterix were reported also by other PIs in 2005. }
\label{table2}
\resizebox{16.5cm}{!}{
\begin{tabular}{lrrrrrrrrr}
\hline
\hline
\noalign{\smallskip}
\noalign{\smallskip}
Asteroid  & $a$ (AU) & $e$ & $i$ ($\deg$) & $\Omega$ ($\deg$) & $\omega$ ($\deg$) & $M$ ($\deg$) & MOID (AU) &  Obs & $\sigma$ ($\arcsec$)\\
\noalign{\smallskip}
\noalign{\smallskip}
\hline

\noalign{\smallskip}\noalign{\smallskip}
\multicolumn{10}{c}{Extended Arcs at First Opposition (Precoveries):} \\
\noalign{\smallskip}

2005 CG41 * &  1.0007879 &  0.2846587 &  19.09737 &  137.74366 &  233.38300 &   89.72675 &  0.0902 &   8 &  0.98 \\
            &  1.0589380 &  0.3535199 &  25.29019 &  137.82014 &  238.16912 &   75.88581 &  0.1170 &  20 &  0.68 \\
\noalign{\smallskip}
2005 TU45   &  1.9737518 &  0.4959559 &  28.53919 &  120.24291 &   76.86254 &  149.07339 &  0.2669 & 808 &  0.31 \\
            &  1.9737518 &  0.4959559 &  28.53919 &  120.24291 &   76.86254 &  149.07339 &  0.2669 & 812 &  0.31 \\  
\noalign{\smallskip}
2005 MB     &  0.9852658 &  0.7927082 &  41.39802 &   88.66029 &   42.80966 &   53.23488 &  0.0840 &  80 &  0.40 \\
            &  0.9852658 &  0.7927081 &  41.39805 &   88.66032 &   42.80964 &   53.23494 &  0.0840 &  83 &  0.40 \\
\noalign{\smallskip}
2004 XP164  &  2.1784898 &  0.4125130 &  22.64397 &  127.01048 &  310.26942 &  109.80532 &  0.3836 & 104 &  0.52 \\
            &  2.1784898 &  0.4125130 &  22.64397 &  127.01048 &  310.26940 &  109.80534 &  0.3836 & 109 &  0.51 \\
\noalign{\smallskip}
2009 TG10   &  1.9737117 &  0.4239201 &  40.88903 &  210.71935 &   12.08752 &   87.39857 &  0.1389 &  47 &  0.39 \\        
            &  1.9737149 &  0.4239234 &  40.88867 &  210.71941 &   12.08990 &   87.39260 &  0.1389 &  53 &  0.37 \\
\noalign{\smallskip}
2009 NA     &  2.6600451 &  0.5529338 &  10.08911 &  269.50371 &   98.88855 &  206.85431 &  0.2669 & 498 &  0.41 \\        
            &  2.6600088 &  0.5529283 &  10.08907 &  269.50366 &   98.88858 &  206.85853 &  0.2669 & 501 &  0.41 \\
\noalign{\smallskip}
2002 HQ11   &  1.8504197 &  0.5955217 &   6.04556 &  153.36013 &  322.08227 &    4.04716 &  0.0555 & 139 &  0.43 \\
            &  1.8504197 &  0.5955216 &   6.04556 &  153.36013 &  322.08227 &    4.04716 &  0.0555 & 141 &  0.43 \\
\noalign{\smallskip}
2001 XK105  &  2.1303469 &  0.5004268 &   7.63156 &   79.92045 &    6.99267 &  104.77324 &  0.0811 &  25 &  0.61 \\        
            &  2.1304028 &  0.5004399 &   7.63171 &   79.92043 &    6.99257 &  104.72642 &  0.0811 &  27 &  0.63 \\ 
\noalign{\smallskip}
2006 WT1    &  2.4717312 &  0.6011840 &  13.68475 &  244.92276 &  170.58186 &  131.36247 &  0.0039 &  79 &  0.46 \\        
            &  2.4717312 &  0.6011840 &  13.68474 &  244.92276 &  170.58186 &  131.36247 &  0.0039 &  85 &  0.45 \\
\noalign{\smallskip}
2002 TY57   &  1.9221114 &  0.3273811 &   3.45506 &  119.03691 &  259.77000 &  194.64386 &  0.3018 & 134 &  0.42 \\
            &  1.9221114 &  0.3273805 &   3.45507 &  119.03677 &  259.77024 &  194.64380 &  0.3018 & 141 &  0.42 \\
\noalign{\smallskip}
2006 PW     &  1.3813093 &  0.6516800 &  35.87687 &  132.98690 &  325.07629 &   88.37690 &  0.3654 &  56 &  0.40 \\        
            &  1.3813093 &  0.6516801 &  35.87686 &  132.98690 &  325.07629 &   88.37691 &  0.3654 &  58 &  0.40 \\
\noalign{\smallskip}
2009 HW2    &  2.2061742 &  0.5266663 &   2.90017 &  211.42013 &  350.74180 &  320.03523 &  0.0419 &  24 &  0.43 \\        
            &  2.1826121 &  0.5214024 &   2.87655 &  211.42082 &  350.69936 &  325.37582 &  0.0422 &  32 &  0.48 \\
\noalign{\smallskip}
2009 SP171  &  1.3557430 &  0.3559500 &  25.62158 &  223.14972 &  285.17007 &   87.36382 &  0.0613 &  54 &  0.46 \\        
            &  1.3557449 &  0.3559509 &  25.62172 &  223.14955 &  285.17012 &   87.36292 &  0.0613 &  56 &  0.46 \\
\noalign{\smallskip}
2005 BY1    &  3.1548620 &  0.6909590 &  17.02151 &  298.02275 &  282.16258 &   73.64347 &  0.1962 &  36 &  0.51 \\        
            &  3.1548619 &  0.6909590 &  17.02150 &  298.02271 &  282.16260 &   73.64349 &  0.1962 &  48 &  0.51 \\
\noalign{\smallskip}
2009 FU23   &  0.8371147 &  0.2820307 &  13.89992 &   57.82262 &  315.14072 &  121.64228 &  0.0344 & 106 &  0.45 \\        
            &  0.8371171 &  0.2820245 &  13.89955 &   57.82290 &  315.14067 &  121.63625 &  0.0344 & 108 &  0.45 \\
\noalign{\smallskip}
2005 VC2    &  2.7670888 &  0.5871642 &  36.78132 &  222.08193 &  166.90560 &  136.56584 &  0.1599 &  95 &  0.40 \\        
            &  2.7670893 &  0.5871642 &  36.78130 &  222.08193 &  166.90566 &  136.56569 &  0.1599 & 100 &  0.40 \\
\noalign{\smallskip}
2006 KA     &  1.6331499 &  0.5614647 &  31.02218 &  236.08104 &  244.60796 &  323.01311 &  0.0693 &  69 &  0.49 \\        
            &  1.6331500 &  0.5614619 &  31.02231 &  236.08099 &  244.60830 &  323.01310 &  0.0693 &  72 &  0.53 \\
\noalign{\smallskip}
2000 FL10   &  1.4629016 &  0.4268866 &  29.01712 &  186.99620 &  258.80260 &  319.18251 &  0.0815 & 218 &  0.55 \\
            &  1.4629011 &  0.4268877 &  29.01714 &  186.99619 &  258.80242 &  319.18377 &  0.0815 & 221 &  0.55 \\
\noalign{\smallskip}

\hline
\noalign{\smallskip}\noalign{\smallskip}
\multicolumn{10}{c}{Extended Arcs at Last Opposition (Recoveries):} \\
\noalign{\smallskip}

1996 XX14 * &  2.5469021 &  0.6518068 &  10.58772 &  195.71905 &  184.95389 &  289.92264 &  0.1036 &  52 &  0.68 \\         
            &  2.5493643 &  0.6501978 &  10.56587 &  195.57444 &  185.06938 &  281.52481 &  0.0997 &  59 &  0.66 \\
\noalign{\smallskip}
2003 MK4 *  &  1.0803203 &  0.1811634 &  22.30645 &  282.63278 &  109.56148 &  181.78607 &  0.0017 & 218 &  0.57 \\        
            &  1.0804562 &  0.1812333 &  22.31694 &  282.63460 &  109.53443 &  181.29420 &  0.0018 & 225 &  0.57 \\
\noalign{\smallskip}
2002 DH2 *  &  2.0507005 &  0.5417798 &  20.97315 &  345.75458 &  231.59884 &  134.88238 &  0.0686 & 160 &  0.53 \\
            &  2.0506639 &  0.5417709 &  20.97292 &  345.75456 &  231.59892 &  134.91606 &  0.0686 & 165 &  0.53 \\
\noalign{\smallskip}
1994 JX     &  2.7631468 &  0.5726370 &  32.16385 &   52.49705 &  193.44617 &  319.74146 &  0.1798 &  92 &  0.58 \\
            &  2.7631468 &  0.5726371 &  32.16383 &   52.49705 &  193.44625 &  319.74147 &  0.1798 & 100 &  0.59 \\
\noalign{\smallskip}
2004 RS25   &  2.1268859 &  0.4795878 &   6.64886 &  179.03636 &  145.29544 &  160.50845 &  0.1140 & 133 &  0.48 \\
            &  2.1268858 &  0.4795879 &   6.64886 &  179.03636 &  145.29543 &  160.50849 &  0.1140 & 135 &  0.47 \\
\noalign{\smallskip}
1999 WK11   &  2.1342937 &  0.4652704 &   7.46410 &   72.67897 &  220.35355 &   57.77729 &  0.1482 & 187 &  0.54 \\
            &  2.1342937 &  0.4652704 &   7.46409 &   72.67897 &  220.35355 &   57.77729 &  0.1482 & 193 &  0.54 \\
\noalign{\smallskip}
2005 GN59   &  1.6565831 &  0.4678204 &   6.62763 &  219.03615 &  202.90753 &  203.39329 &  0.0501 & 570 &  0.43 \\
            &  1.6565831 &  0.4678204 &   6.62763 &  219.03615 &  202.90753 &  203.39329 &  0.0501 & 572 &  0.43 \\
\noalign{\smallskip}
2002 EQ9    &  1.8356544 &  0.4651654 &  16.30670 &  179.23417 &   44.14087 &  347.65988 &  0.0626 & 362 &  0.45 \\
            &  1.8356544 &  0.4651654 &  16.30670 &  179.23417 &   44.14086 &  347.65988 &  0.0626 & 364 &  0.45 \\
\noalign{\smallskip}
2007 DK     &  1.3961987 &  0.5503815 &   5.17549 &  290.95892 &  354.89948 &  311.16457 &  0.0894 &  85 &  0.55 \\
            &  1.3961987 &  0.5503811 &   5.17550 &  290.95910 &  354.89925 &  311.16454 &  0.0894 &  89 &  0.55 \\
\noalign{\smallskip}
2005 CA     &  2.7290438 &  0.5890431 &  16.75355 &  202.13893 &  203.97095 &  242.81789 &  0.1495 &  82 &  0.60 \\
            &  2.7290438 &  0.5890431 &  16.75355 &  202.13891 &  203.97096 &  242.81791 &  0.1495 &  86 &  0.59 \\
\noalign{\smallskip}

\hline
\hline
\end{tabular}
}
\end{center}
\end{table*}

\begin{table*}[!t]
\begin{center}
\caption{28 instrument archives available in August 2012 in the {\it Mega-Archive} used by {\it Mega-Precovery} 
adding together about 2.5 million images. We list the telescope, instrument, number of images (thousands rounded), archive 
start and end date, field of view (in arcmin), number of CCDs (for mosaics) and estimated $V$ limiting magnitude suitable to 
detect NEAs. } 
\label{table3}
\begin{tabular}{llrrrrrr}
\noalign{\smallskip}
\noalign{\smallskip}
\hline
\hline
\noalign{\smallskip}
\noalign{\smallskip}
Telescope  & Instrument & Nr. images & Start Date & End  Date  & FOV $^{\prime}$ & CCDs & $V$ \\
\noalign{\smallskip}
\noalign{\smallskip}
\hline
\hline
\noalign{\smallskip}\noalign{\smallskip}

\multicolumn{8}{c}{ESO Instruments: } \\
\noalign{\smallskip}\noalign{\smallskip}

VLT 8.2m   &  FORS1        &   36,000  &  23-01-1999  &  26-03-2009  &    $6.8\times6.8$ &  2 & 26 \\
VLT 8.2m   &  FORS2        &  111,000  &  30-10-1999  &  25-02-2012  &    $6.8\times6.8$ &  2 & 26 \\
VLT 8.2m   &  HAWKI        &   69.000  &  01-08-2007  &  24-02-2012  &    $7.5\times7.5$ &  4 & 26 \\
VLT 8.2m   &  ISAAC        &  199,000  &  01-03-1999  &  25-02-2012  &    $2.5\times2.5$ &  1 & 26 \\
VLT 8.2m   &  NACO         &  275,000  &  02-12-2001  &  29-02-2012  &    $1.0\times1.0$ &  1 & 26 \\
VLT 8.2m   &  VIMOS        &   66,000  &  30-10-2002  &  28-02-2012  &  $12.8\times16.0$ &  4 & 26 \\
VLT 8.2m   &  VISIR        &   67,000  &  11-05-2004  &  26-02-2012  &    $0.5\times0.5$ &  1 & 26 \\
VISTA 4.1m &  VIRCAM       &  230,000  &  16-10-2009  &  22-06-2011  &  $46.3\times46.3$ & 16 & 25 \\
VST 2.6m   &  OmegaCam     &   19,000  &  01-04-2011  &  15-03-2012  &  $58.4\times58.4$ & 32 & 24 \\
NTT 3.5m   &  EMMI         &   18,000  &  17-03-2004  &  01-04-2008  &    $9.1\times9.1$ &  2 & 25 \\
NTT 3.5m   &  SOFI         &  126,000  &  30-03-2006  &  15-02-2012  &    $4.9\times4.9$ &  1 & 25 \\
NTT 3.5m   &  SUSI2        &   17,000  &  02-04-2004  &  29-12-2008  &    $5.5\times5.5$ &  2 & 25 \\
ESO 3.6m   &  EFOSC2       &   47,000  &  03-07-2004  &  16-03-2012  &    $4.1\times4.1$ &  1 & 25 \\
ESO 3.6m   &  TIMMI2       &   64,000  &  08-05-2004  &  28-06-2006  &    $1.6\times1.2$ &  1 & 25 \\
ESO/MPG 2.2m &  WFC        &  124,000  &  20-06-1998  &  25-02-2012  &  $33.6\times32.7$ &  8 & 23 \\
\noalign{\smallskip}\noalign{\smallskip}

\hline
\noalign{\smallskip}\noalign{\smallskip}
\multicolumn{8}{c}{AURA NVO Instruments: } \\
\noalign{\smallskip}\noalign{\smallskip}

KPNO 4m   &  MOSAIC      &   33,000  &  01-09-2004  &  27-06-2012  &      $36\times36$ &  8 & 25 \\
KPNO 4m   &  NEWFIRM     &  130,000  &  30-06-2007  &  10-07-2012  &      $28\times28$ &  4 & 25 \\
WIYN 3.5m &  Mini Mosaic &    6,000  &  17-03-2009  &  19-07-2012  &      $10\times10$ &  2 & 25 \\
WIYN 3.5m &  WHIRC       &   89,000  &  04-04-2009  &  11-04-2012  &    $3.3\times3.3$ &  1 & 25 \\
WIYN 0.9m &  MOSAIC      &    9,000  &  27-05-2009  &  03-05-2012  &      $59\times59$ &  8 & 21 \\
CTIO 4m   &  MOSAIC-2    &   67,000  &  11-08-2004  &  20-02-2012  &  $37.0\times37.5$ &  8 & 25 \\
CTIO 4m   &  NEWFIRM     &   74,000  &  18-05-2010  &  17-10-2011  &      $28\times28$ &  4 & 25 \\
CTIO 0.9m &  Cass Img    &  228,000  &  27-03-2009  &  24-07-2012  &  $13.5\times13.5$ &  1 & 21 \\
SOAR 4m   &  OSIRIS      &   60,000  &  17-03-2009  &  20-07-2012  &    $3.3\times3.3$ &  2 & 25 \\
\noalign{\smallskip}\noalign{\smallskip}

\hline
\noalign{\smallskip}\noalign{\smallskip}
\multicolumn{8}{c}{Other Instruments: } \\
\noalign{\smallskip}

CFHT 3.6m   &  CFHTLS       &   25,000  &  22-03-2003  &  02-02-2009  &  $57.6\times56.4$ & 36 & 25 \\
INT 2.5m    &  WFC          &  230,000  &  20-06-1998  &  10-07-2009  &  $34.1\times34.5$ &  4 & 23 \\
Subaru 8.3m &  SuprimeCam   &   60,000  &  05-01-1999  &  31-12-2010  &  $35.1\times27.6$ & 10 & 26 \\
AAT 3.9m    &  WFC          &    5,000  &  21-08-2000  &  05-02-2006  &  $31.4\times31.4$ &  8 & 25 \\

\noalign{\smallskip}
\noalign{\smallskip}
\hline
\hline

\end{tabular}
\end{center}
\end{table*}


\begin{thebibliography}{}
   \bibitem[Baade et al., 1999]{baa99} Baade, D.; Meisenheimer, K.; Iwert, O.: 1999, {\it The Messenger} 95, 15
   \bibitem[Berthier et al., 2006]{ber06} Berthier, J., Vachier, F., Thuillot, W., Fernique, P., Ochsenbein, F., 
            Genova, F., Lainey, V., Arlot, J.-E.: 2006, {\it Astronomical Data Analysis Software and Systems} XV, 351, 367
   \bibitem[Berthier et al., 2009]{ber09} Berthier, J., Hestroffer, D., Carry, B., Vachier, F., Lainey, V., Emelyanov, N.~V., Thuillot, W., Arlot, J.-E.: 2009, 
           {\it Miriade: A Service for Solar Sytem Ojects Ephemerides in the VO Framework}, in the {\it European Planetary Science Congress 2009}, p. 676 (Sep 2009)
   \bibitem[Boattini et al., 2004]{boa04} Boattini, A., et al., 2004, Astronomy \& Astrophysics, 418, 743
   \bibitem[Bout, 2007]{bou07} Bout, J.: 2007, {\it Searching for known asteroids in the WFI archive using Astro-WISE}, MSc project 
           (supervisors: Kleijn G. V. and Valentijn, E.), http://www.astro.rug.nl/\~bout/small-research/small-research-asteroids.pdf
   \bibitem[Drew et al., 2005]{dre05} Drew, J., et al.: 2005, \mnras 362, 753
   \bibitem[Elst et al., 2011]{els11} Elst, E., Boattini, A., Behrend, R., Vaduvescu, O., Galad, A., Popescu, M., Comsa, I., Paraschiv, A., 
            Lacatus, D., Suciu, O., Sonka, A., Tudorica, A., Badescu, T., Badea, M., Constantinescu, M., Vidican, D., Opriseanu, C.: 2011, MPC 77265
   \bibitem[EURONEAR, 2012]{eur12} EURONEAR, 2012 - {\it Observing Tools - Archive Mega-Precovery}, http://euronear.imcce.fr/tiki-index.php?page=MegaPrecovery
   \bibitem[Fitzsimmons et al., 2011]{fit11} Fitzsimmons, A., Balam, D., Vaduvescu, O., Licandro, J., Patrick, L., Karami, M., Sonka, A., Popescu, M.,
            Comsa, I., Paraschiv, A., Lacatus, D., Suciu, O., Tudorica, A., Badescu, T., Badea, M.. Constantinescu, M., Vidican, D., Opriseanu, C.: 2011, MPC 77266
   \bibitem[Gray, 2012]{gra12} Gray, B., 2012, FIND\_ORB software, http://www.projectpluto.com/find\_orb.htm
   \bibitem[Groot et al., 2009]{gro09} Groot, A., et al., 2009, Monthly Notices of the Royal Astronomical Society 399, 323
   \bibitem[Gwyn et al., 2011]{gwy11} Gwyn, S., Hill, N. \& Kavelaars, J. J.: 2011, {\it SSOS: A Moving Object Image Search Tool for Asteroid Precovery at the CADC}, 
           http://arxiv.org/abs/1111.3364, communication ADASS XXI (2011)
   \bibitem[IMCCE, 2012a]{imc12a} IMCCE, 2012a, {\it The IMCCE Virtual Observatory Solar System Portal}, http://vo.imcce.fr/webservices/miriade
   \bibitem[IMCCE, 2012b]{imc12b} IMCCE 2012b, {\it Connaissance des Temps}, cap. 8
   \bibitem[Ivezic, 2008]{ive08} Ivezic, Z.: 2008, {\it The 4th Release of the Sloan Digital Sky Survey Moving Object Catalog}, 
           http://www.astro.washington.edu/users/ivezic/sdssmoc/sdssmoc.html
   \bibitem[Kent et al., 2009]{ken09} Kent, S. M., et al., 2009 {\it Survey of Near-Earth Objects from the SDSS}, AAS Meeting 214, 435.05; Bulletin of the AAS, 41, 707
   \bibitem[Popescu et al., 2010]{pvb10} Popescu, M., Vaduvescu, O. \& Birlan, M.: 2010, 
           {\it Mega-Precovery, a dedicated project for data mining worldwide image archives for poorly known asteroids}, 
           Journees Scientifiques 2010 de IMCCE, Paris, http://euronear.imcce.fr/tiki-download\_file.php?fileId=474
   \bibitem[Popescu and Vaduvescu, 2010]{pv10} Popescu, M. and Vaduvescu, O.: 2010, 
           {\it Mega-Precovery, a dedicated project for data mining worldwide image archives for poorly known asteroids}, 
            Romanian Diaspora Conference, http://www.diaspora-stiintifica.ro/diaspora2010/prezentare18.html
   \bibitem[Popescu and Vaduvescu, 2011]{pv11} Popescu, M. and Vaduvescu, O.: 2011, 
           {\it EURONEAR website: Observing Tools - Archive Mega Precovery}, http://euronear.imcce.fr/tiki-index.php?page=MegaPrecovery
   \bibitem[Raab, 2012]{raa12} Raab, H.: 2012, {\it Astrometrica software}, http://www.astrometrica.at
   \bibitem[Skvarc, 2012]{skv12} Skvarc, J, 2012, {\it Calculation of residuals of asteroid positions} http://www.fitsblink.net/residuals
   \bibitem[Solano et al., 2011]{sol11a} Solano, E., Rodrigo, C, Vaduvescu, O.: 2011,
            {\it PHA's precovery using the Virtual Observatory. Solar System Exploration}, communication (in Spanish), 
            IVOA Executive Committee Meeting (FM42), Bilbao. June 2011
   \bibitem[Solano, 2011]{sol11b} Solano, E.: 2011, {\it private communication}
   \bibitem[Spahr \& Williams, 2011]{spa11} Spahr, T. \& Williams, G.: 2011, {\it private communication}
   \bibitem[Steel, 1997]{ste97} Steel, D., et al.: 1997 Australian Journal of Astronomy, 7, 67
   \bibitem[SVO, 2011]{svo11} Spanish Virtual Observatory (SVO): 2011, {\it Identificacion de asteroides cercanos a la Tierra} (in Spanish), 
           http://www.laeff.cab.inta-csic.es/projects/near/main
   \bibitem[Tristan, 2011]{mun11} Tristan, R. M.: 2011, {\it Descubra desde su ordenador un asteroide peligroso para la Tierra}, 
           {\it El Mundo} magazine (in Spanish), http://www.elmundo.es/elmundo/2011/07/21/ciencia/1311266218.html
   \bibitem[Vaduvescu et al., 2009]{vad09} Vaduvescu, O. et al: 2009, AN 330, 7, 698
   \bibitem[Vaduvescu et al., 2011a]{vad11a} Vaduvescu, O. et al: 2011a, AN 332, 6, 580
   \bibitem[Vaduvescu et al., 2011b]{vad11b} Vaduvescu, O. et al: 2011b, P\&SS 59, 1632
   \bibitem[Yau et al., 2011]{yau11} Yau, K. K., et al: 2011, {\it MOST - Moving Object Search Tool for NEOWISE and IRSA}, AAS Meeting 217, 333.18; 
            Bulletin of the American Astronomical Society, Vol. 43

\end{thebibliography}
\end{document}